\documentclass[conference]{IEEEtran}
\IEEEoverridecommandlockouts

\ifCLASSINFOpdf
  \usepackage[pdftex]{graphicx}
\else
\fi

\usepackage{url}
\usepackage{okumoto}
\usepackage{array}
\usepackage{multirow}
\usepackage{newtxtext,newtxmath}

\newcolumntype{P}[1]{>{\centering\arraybackslash}p{#1}}
\newcommand{\figref}[2][]{Fig.~\ref{#2}}

\newif\ifblind

\begin{document}

\title{Response Collector:\\ A Video Learning System for Flipped Classrooms}

\ifblind
\else
  \author{\IEEEauthorblockN{Hayato Okumoto, Mitsuo Yoshida and Kyoji Umemura}
  \IEEEauthorblockA{Department of Computer Science and Engineering\\
  Toyohashi University of Technology \\
  Toyohashi, Aichi, Japan\\
  h153317@edu.tut.ac.jp, yoshida@cs.tut.ac.jp, umemura@tut.jp
  }
  \and
  \IEEEauthorblockN{Yuko Ichikawa}
  \IEEEauthorblockA{General Education Department\\
  National Institute of Technology,
  Tokyo College\\
  Hachioji, Tokyo, Japan\\
  yuko@tokyo-ct.ac.jp}
  }
\fi

\maketitle

\begin{abstract}
  The flipped classroom has become famous as an effective educational method that flips the purpose of classroom study and homework.
  In this paper, we propose a video learning system for flipped classrooms, called Response Collector, which enables students to record their responses to preparation videos.
  Our system provides response visualization for teachers and students to understand what they have acquired and questioned.
  We performed a practical user study of our system in a flipped classroom setup.
  The results show that students preferred to use the proposed method as the inputting method, rather than naive methods.
  Moreover, sharing responses among students was helpful for resolving individual students' questions, and students were satisfied with the use of our system.
\end{abstract}


\section{Introduction}
Recently, the flipped classroom has been recognized by educators as a notable teaching method~\cite{Bergmann2012,Bishop2013,Thai2017}.
The flipped classroom is an active learning method
in which students watch videos to absorb basic knowledge before class hours.
During class hours, they mostly use this knowledge on problem-solving exercises and discussions.
This form of lecture enables students to learn at their own pace by skipping and rewinding videos during their preparation time~\cite{Thai2017}.
Moreover, teachers and students can take more time to communicate and interact with each other in the classroom.
Thus, the flipped classroom turns lectures into student-centered learning.

In the flipped classroom, it is important that teachers can understand what students have absorbed during preparation.
If teachers can anticipate students' questions and what they have not understood during preparation, they can provide supplementary explanations during class hours.
As an uncomplicated example, Bergmann and Sams had their students take notes and record any questions they had.
They then discussed these questions at the beginning of the class~\cite{Bergmann2012}.

\IEEEpubidadjcol

In this paper, we propose a video learning system for flipped classrooms, called Response Collector, which enables students to record their responses to preparation videos.
This system aims at improving lectures by collecting responses from students and providing visualized responses to teachers and students.
This paper focuses on answering the following research questions:

\begin{LaTeXdescription}
  \item[RQ (A)] Can Response Collector collect more responses than naive methods?
  \item[RQ (B)] Which type of inputting method do students prefer?
  \item[RQ (C)] Is it useful for students to share information about their responses in the classroom?
  \item[RQ (D)] Are students satisfied with flipped classes where the teacher gives feedback on their responses from Response Collector?
\end{LaTeXdescription}

\IEEEpubidadjcol

\fig[width=\columnwidth]{The user interface of Response Collector.
There are some basic functions to control the player: playing, pausing, seeking, skipping 10 seconds, and adjusting play speed.
There are also four buttons to help respond to the video, and these responses are visualized.
Teachers and students can instantaneously assess instances where students responded a lot.
}{fig_interface}{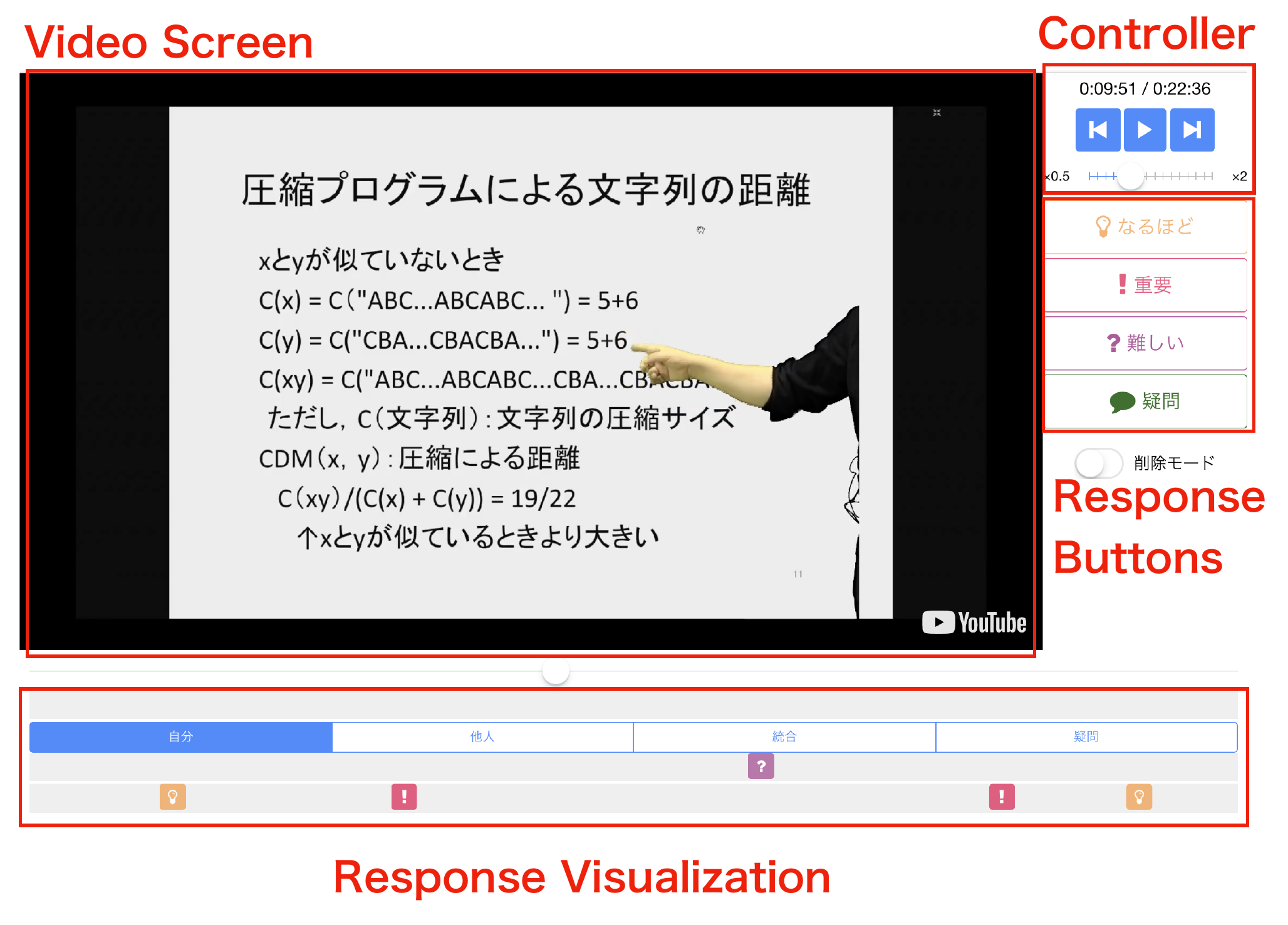}

\section{Related Work}
\subsection{Video Annotation}
Video annotation systems for educational use have been developed by~\cite{Bargeron1999,Bargeron2002,Chatti2016,Risko2013}.
There are two types of annotations: free text annotations and fixed categorical annotations.

For free text annotations, Bargeron et al. have developed the Microsoft Research Annotation System (MRAS)~\cite{Bargeron1999,Bargeron2002}.
The MRAS has functions which can record typed text annotations and spoken audio annotations.
Bargeron et al. compared the MRAS with the basic pen and paper method, in terms of being an inputting method, and their results demonstrated that the former was preferred over the latter.
They also showed that the number of comments and questions increased when they shared annotations among participants.
A recent similar system is CourseMapper, which focuses on video annotation and annotation visualization for video-based learning environments~\cite{Chatti2016}.
However, this visualization does not distinguish types of annotations. Therefore, it is difficult to understand at a glance where each type of annotation peaks.
These studies, however, only focused on video annotations
and did not extract or classify intentions from text annotations.
Thus, we need to know what parts were confusing to students, what parts they were interested in, and what parts they had questions about.
For this reason, free text annotation is not applicable for our purposes.

As for fixed categorical annotations, Risko et al. developed the Collaborative Lecture Annotation System (CLAS)~\cite{Risko2013},
which has only one category of annotation.
Using the CLAS, each student indicates important points through a simple button press.
The CLAS then visualizes the points of interest of all students in the videos.
We adopted this concept for our system.

Note that these related studies did not focus on the flipped classroom.
We concluded that video annotation would work well with preparation videos for flipped classrooms because students and teachers could then make use of the response information during class hours.

\fig[width=\columnwidth]{Some ``question'' responses are shown in this list view.
(English translations have been added to facilitate understanding.)
By clicking on each row, users can jump to the position where the question was recorded.}{fig_questions}{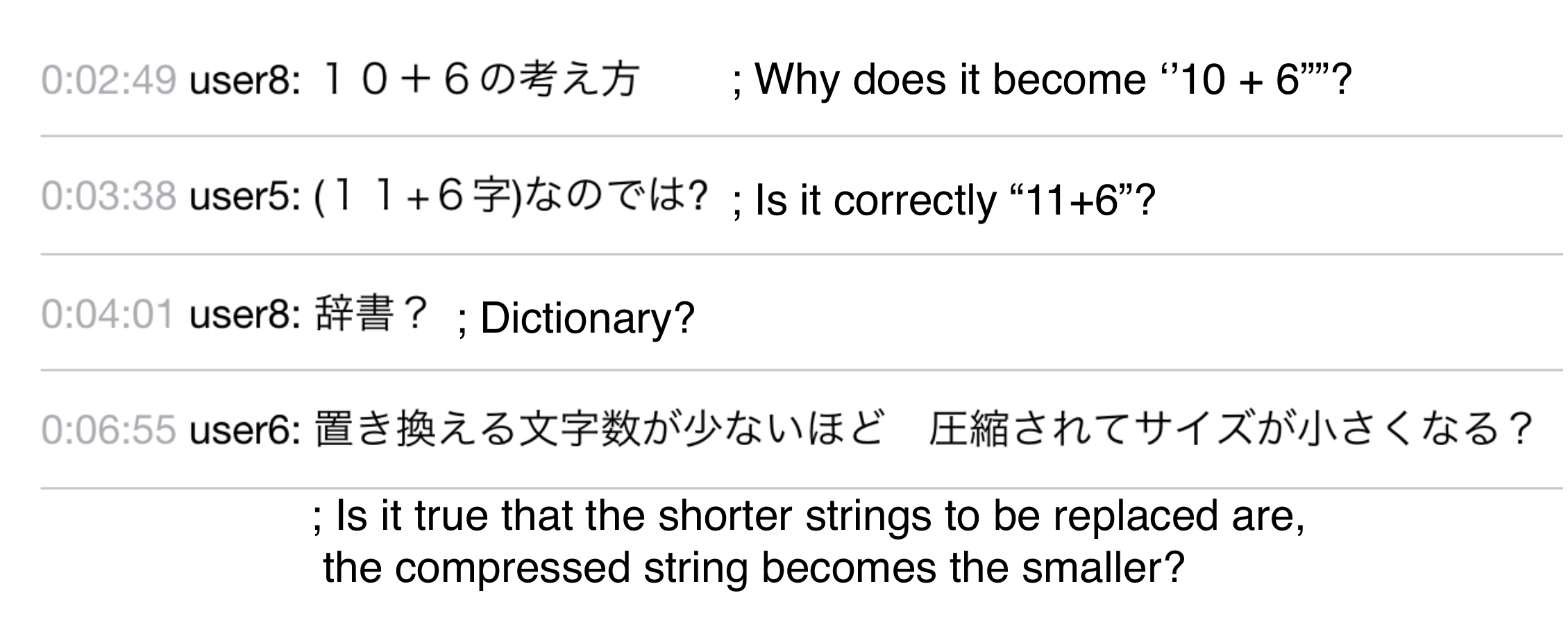}

\subsection{Learning Analysis of User Behavior}
A number of works have aimed at uncovering student profiles from user interaction logs (such as player control events: Play, Pause, Fast-forward, Rewind, etc.) on video learning systems~\cite{Mirriahi2016,Kim2014,Shi2015}.
Mirriahi et al. aimed to uncover students' learning profiles from the behavior logs of their video annotation tool~\cite{Mirriahi2016},
and they classified user profiles into four clusters.
This study, however, was not concerned with specific video analysis.
Kim et al. have analyzed user interaction peaks on Massive Open Online Course (MOOC) videos~\cite{Kim2014}.
Their analysis demonstrated that user interaction peaks might indicate areas of user confusion or points of interest in the video material.
Shi et al.~\cite{Shi2015} developed VisMOOC, which visualizes user clickstream of MOOC platforms to help analyze user learning behaviors.
It is interesting to uncover user behavior by only analyzing the clickstream.
However, these works~\cite{Kim2014,Shi2015} require large amounts of data, such as clickstream logs on MOOC platforms,
and therefore, may not be applicable for classroom use.

\fig[width=\columnwidth]{This figure shows the visualization of responses,
separated into the responses of each student.
By clicking each response icon, users can jump to the position where the response is recorded.}{fig_separate}{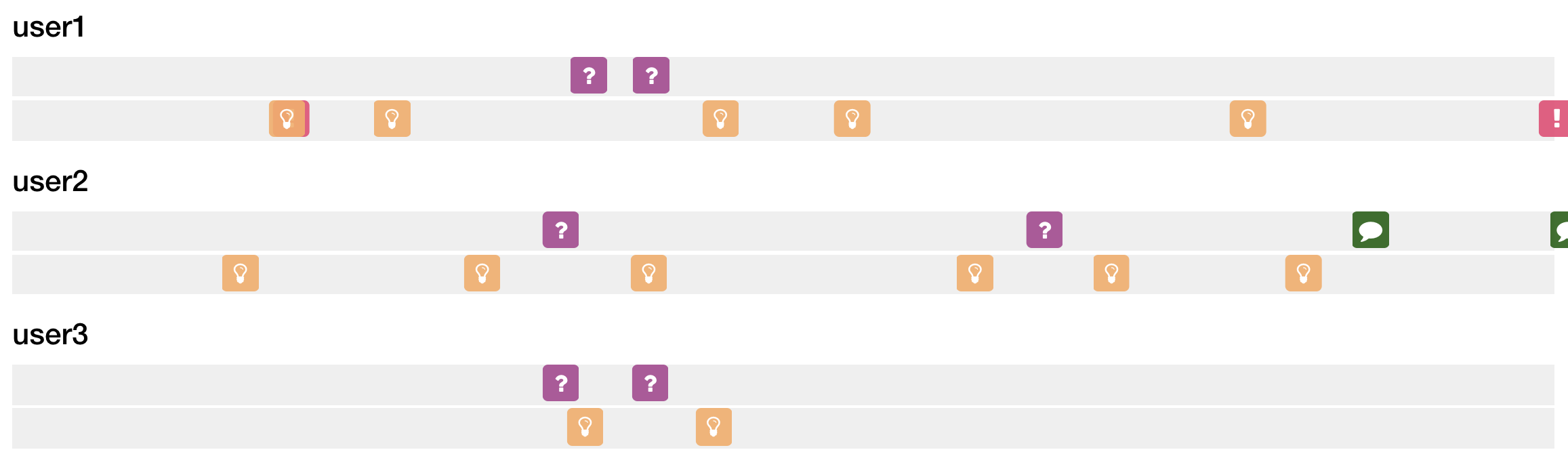}
\fig[width=\columnwidth]{This figure shows the aggregated visualization of responses.
Each line corresponds to a different response type.
The height of the line corresponds to the total number of responses at a position.
}{fig_aggregation}{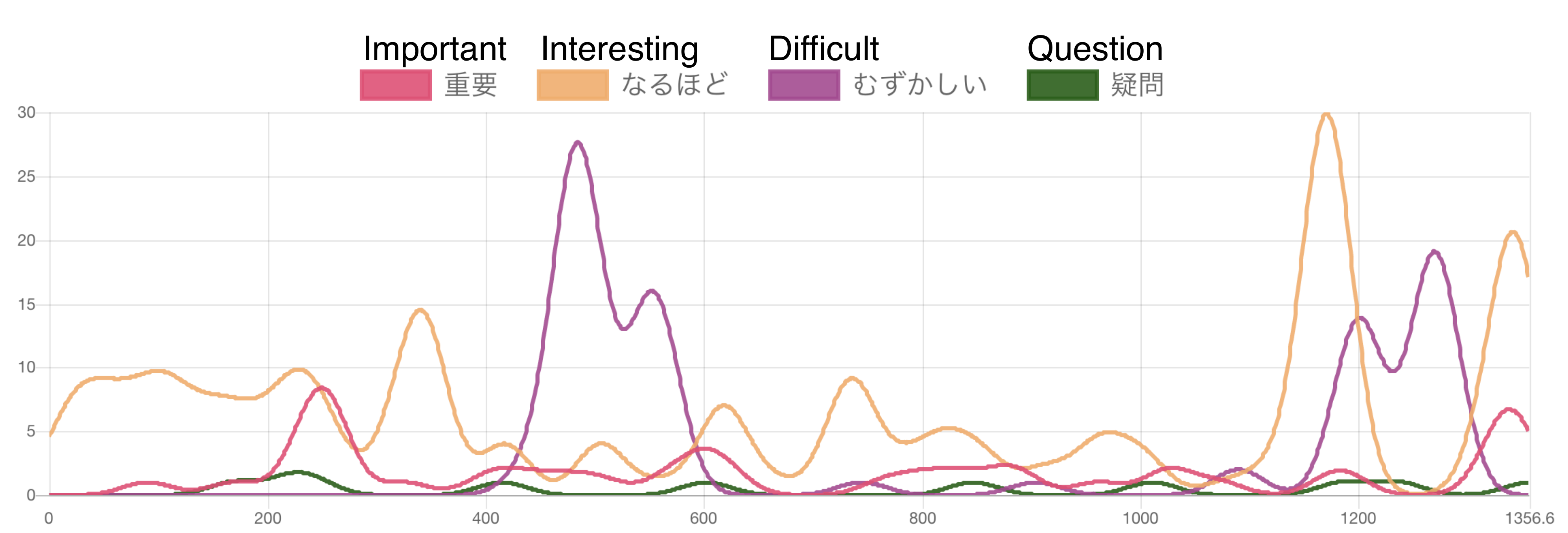}

\section{System Design}
We developed Response Collector as a combination video annotation and annotation visualization system for preparation videos in flipped classrooms.
We adopted fixed categorical annotations for Response Collector because of their simplicity in terms of inputting by clicking a button.
We also needed to visualize whether or not each type of response peaks.
In Response Collector, ``annotation'' corresponds to ``response.''

\subsection{Types of Response}
We pondered on the types of responses in the system design
and defined them as follows:
\begin{itemize}
  \item Interesting
  \item Important
  \item Difficult
  \item Question (free text)
\end{itemize}
We assumed that the length of the video would be approximately 10 to 20 minutes
and on that basis defined four types so that students would not be puzzled about which type of response to use.

For ``Interesting'' and ``Important'' responses, we intended for students to mark interesting and important parts of the video respectively.
This allowed them the opportunity to expand on the details of content which they responded to as interesting.
We believe that these responses can be used not only for lectures, but also for reviewing videos efficiently.

During class hours, we assumed that an important role for teachers would be to resolve areas that students found difficult and areas that they had questioned.
For this purpose, we defined the types ``Difficult'' and ``Question.''
If the teachers were to explain those areas marked as ``Difficult'' or ``Question'' during class hours, it would lead to efficient teaching and student satisfaction.

\subsection{Features}
We implemented Response Collector as a Web application that could be simply used by students without installing any additional software.
Response Collector identifies students by their login ID and password.
\figref{fig_interface} shows the user interface of the Response Collector video player.
This player has basic player functions (playing, pausing, seeking, skipping 10 seconds, and adjusting play speed).

The characteristic features of Response Collector are the response input buttons and response visualization.
Students can record a response by simply clicking a response button.
Response Collector records the response at the current playback position.
These responses are immediately sent from the web browser to a data storage server.
Only when the ``question'' button is clicked, the video is temporarily paused and a text dialog box appears, letting students type their questions.
The ``question'' responses are shown in list view (\figref{fig_questions}).

The server responds with the aggregation data of all responses.
The web browser then renders visualization graphs from them.
\figref{fig_separate}, \ref{fig_aggregation} shows the response visualizations.
\figref{fig_separate} comprises the separated visualization of each student.
\figref{fig_aggregation} comprises the aggregated visualization of all.
Each line corresponds to a type of response.
The height of the line corresponds to the total number of responses at a position.
From the peak of this graph, teachers and students can identify the position with the highest number of responses for each response type in the video.

\subsection{Response Collector in the Flipped Classroom}
In this section, we introduce an example use of Response Collector in a flipped classroom.
Before class hours, students use Response Collector to watch preparation videos and input responses.
At the beginning of the class, teachers provide supplementary feedback for students' recorded responses.
The peaks of the aggregation graph represent consensus among students.
Therefore, teachers can mainly focus on imparting knowledge about those areas where the peaks occur.
In addition, teachers can answer the questions that students have recorded.
We consider it important to answer and discuss these questions because other students may come up with more questions or ideas through the discussion.
After the feedback sessions, students begin their active learning time by participating in problem-solving exercises or group discussions.
After class, students can review the contents of the lecture using Response Collector.
Students mainly focus on reviewing ``important'' responses.
This will comprise a digest of the video and lead to reduced reviewing time.

\section{User Study}
We performed a practical user study of our proposed system.
We introduced our system to the classroom environment, and students used it to watch a video and respond accordingly.
We then carried out two participant surveys.
In this user study, we provided preparation time at the beginning of each lecture.
We will discuss this in section \ref{sec_discussion}.

In this study, we formulated the following research questions:

\begin{LaTeXdescription}
  \item[RQ (A)] Can Response Collector collect more responses than naive methods?
  \item[RQ (B)] Which type of inputting method do students prefer?
  \item[RQ (C)] Is it useful for students to share information about their responses in the classroom?
  \item[RQ (D)] Are students satisfied with flipped classes where the teacher gives feedback on their responses from Response Collector?
\end{LaTeXdescription}

\subsection{Environments}
To perform this user study, we selected a computer programming course from the many courses in our department.
Flipped classrooms requires students to prepare for class and then mostly focus on doing exercises during class.
In our selected course, students mainly practice programming.
At the beginning of the class, teachers provide necessary knowledge and the details of the exercise.
Students then begin the exercise.
It would be the best course to examine the effectiveness of our system.

\figref{fig_view_of_classroom} shows the classroom environment.
This room is designed for a class where students need to use a computer.
Students can use these computers or their own laptops.
There is also a projection screen and a whiteboard.
In this environment, teachers can give lectures and supplement explanations, using these screen and whiteboard seamlessly.

\fig[width=0.98\columnwidth]{This picture shows the environment of the classroom where we conducted the user study. In this classroom, there are computers connected to the Internet for each student using the Response Collector to browse videos. There is also a projection screen and a whiteboard. Teachers can provide comments, just like in a lecture.}{fig_view_of_classroom}{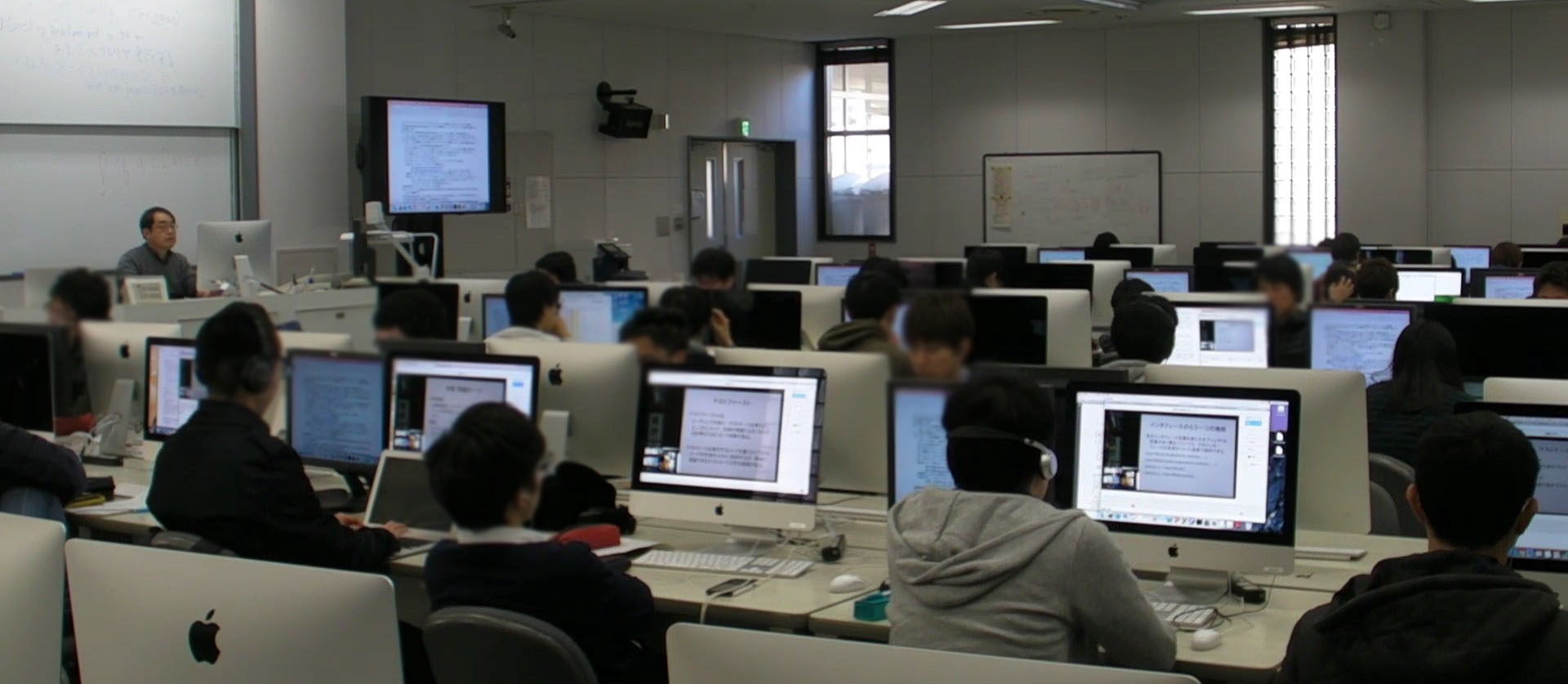}

\begin{table*}[t]
\caption{Usage Statistics of Each Video}
\label{tab_usage_statistics}
\fontsize{7pt}{6.6pt}\selectfont

\begin{center}
\begin{tabular}{P{4mm}|P{13mm}|cP{30mm}c|cc|P{9mm}P{9mm}P{9mm}P{9mm}|P{4mm}P{4mm}P{4mm}}
 & & \multicolumn{3}{c|}{Video} & \multicolumn{2}{c|}{Number of Students} & \multicolumn{7}{c}{Number of Responses} \\
Lect. & Method & \# & Title & Length & Played & Respond & Interesting & Important & Difficult & Question & Sum. & Ave. & Std. \\\hline
I  & Pen \& Paper & 1 & Introduction to Git & 22:42 & (42) & 34 & 46 & 51 & 10 & 20 & 127 & 3.0 & 3.0 \\\hline
   & Google Spreadsheet & 2 & Git with GitHub & 15:27 & (42) & 36 & 43 & 22 & 2 & 5 & 72 & 1.7 & 1.4 \\\hline
II & \multirow{9}{*}{\shortstack{Response\\Collector}} & 3 & Interface Specification and Prototyping & 11:57 & 42 & 40 & 101 & 47 & 8 & 4 & 160 & 3.8 & 2.2 \\
III &&& No Video &&&  &  &  &  &  &  &  &  \\
   && 4 & Software Test & 12:32 & 41 & 38 & 86 & 71 & 5 & 5 & 167 & 4.1 &2.9 \\
IV && 5 & Pair Programming & 11:07 & 42 & 36 & 69 & 86 & 3 & 0 & 158 & 3.8 & 3.4 \\
   && 6 & Suffix Array & 06:21 & 42 & 33 & 47 & 22 & 9 & 2 & 80 & 1.9 & 1.6 \\
V  && 7 & Refactoring & 09:47 & 41 & 34 & 69 & 50 & 1 & 0 & 120 & 2.9 & 2.5 \\
   && 8 & Dynamic Programming & 13:49 & 42 & 33 & 66 & 49 & 28 & 0 & 143 & 3.4 & 3.4 \\
VI && 9 & Code Reading & 07:33 & 34 & 23 & 61 & 12 & 0 & 0 & 73 & 2.1  2.3 \\
&& 10 & About Inheritance & 12:47 & 35 & 24 & 64 & 39 & 5 & 0 & 108 & 3.1 & 3.9
\end{tabular}
\end{center}
\end{table*}

\subsection{Participants}
We asked the students from the computer programming course for their cooperation in this user study, and 42 out of 43 students agreed to participate.
They were all undergraduate students of approximately 21 years of age.

\subsection{Methods}
Our interest lay in the question: ``Can Response Collector collect more responses than naive methods?''
We compared other methods that can collect students' responses.
Note that we only paid attention to comparing the difference in the input interface.
We compared Response Collector with the ``pen and paper'' method as used in Bergeron et al. ~\cite{Bargeron1999},
as well as with a collaborative web-based system ``Google Spreadsheets\footnote{Google Spreadsheets: \url{https://docs.google.com/spreadsheets/}}''
which enabled us to share data in real-time.

Students wrote the following information on sheets of paper or typed it onto spreadsheets:
\begin{itemize}
  \item Time on Video
  \item Type of Response
  \item Content of Question
\end{itemize}
The difference between these methods and Response Collector is that for the former types, students needed to write or type the time of the video manually.
Furthermore, these responses were not visualized into a graph like the one Response Collector generates.
During the feedback session, the teacher solely answered questions that students wrote or typed.

\subsection{Procedure}
We performed this user study during the regular course,
which comprised three hours of class time per week and was held six times per semester.
We had students watch videos using Response Collector.

The procedure of each class was as follows.
At the beginning of the class, students watched two videos and input responses into Response Collector for approximately 30 minutes.
The teacher then gave feedback on their responses.
The teacher mainly explained the content for which the aggregation graph peaked and also answered questions.
After the feedback, students began an exercise.
Students could use Response Collector both during the exercise and after class.
We also enabled them to use the system outside the university, so that they could use it wherever an Internet connection was available.

In the first class, we had students use the ``pen and paper'' and Google Spreadsheets methods to compare the difference between Response Collector and them.
In the final class, we had students watch videos before class to clarify how many students would not watch the video if they were asked to watch in advance.

We conducted surveys of the participants after they had used all methods just one time (at the second lecture) and after the whole course had finished (at the sixth lecture).

\section{Results}

\subsection{Usage Statistics}
\fig[width=0.98\columnwidth]{This figure shows the normalized average number of responses on each video. Students responded to video \#1 using pen and paper and video \#2 using Google Spreadsheets. They watched videos from \#3 to \#10 using Response Collector. The number of responses collected by Response Collector was more than twice as many as obtained by the other methods.}{fig_norm_responses}{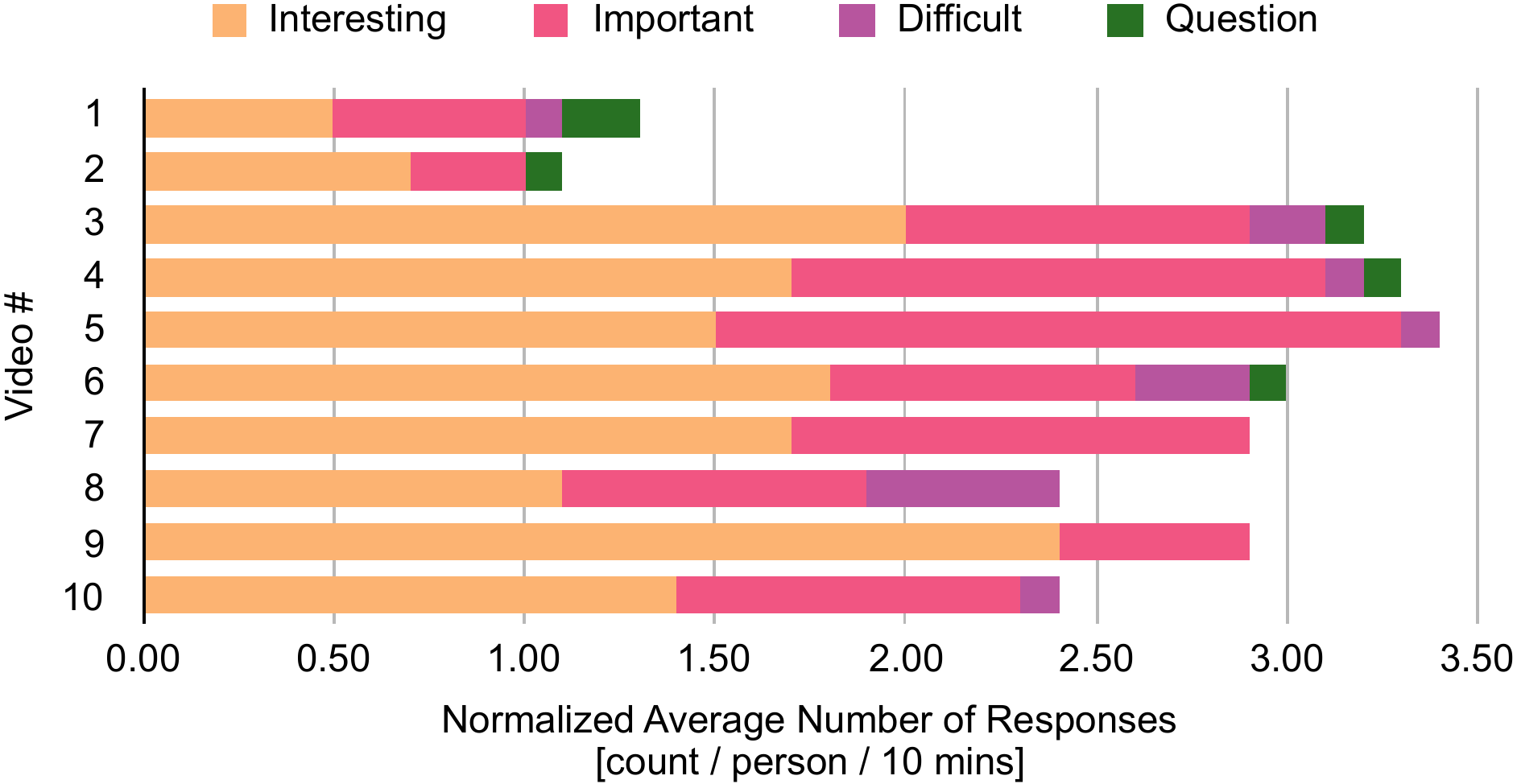}

This section answers RQ(A): ``Can Response Collector collect more responses than naive methods?''

Table~\ref{tab_usage_statistics} shows the usage statistics of each video.
Response Collector can record play histories that correspond to each section that the student watched.
We counted the number of play counts if the student had watched for at least one second.
This number does not represent the count before each class had begun but the total number of play counts.

First, we examined the number of students who played the video.
The results show that most students watched videos from \#3 to \#8.
We had students play videos \#9 and \#10 before class.
Remarkably, the number of play counts for \#9 and \#10 were less than that for videos \#3 to \#8.
This indicates that some students did not watch the video when instructed to view it in advance.

We then examined the number of responses inputted by students.
The sum of responses varied between videos.
We found a positive correlation between the length of the video and the average number of responses ($R^2 = 0.74$).
Thus, we normalized the average number of responses with the length of the video.
\figref{fig_norm_responses} shows the normalized average number of responses in each video.
Using pen and paper, students input 1.3 responses per person per 10 minutes.
They input 1.1 responses on average using Google Spreadsheets.
In contrast, using Response Collector, they input 3.0 responses on average.
These data show that the number of responses collected by Response Collector was more than twice as many as the number of responses obtained by the pen and paper or Google Spreadsheets methods.

\fig[width=0.98\columnwidth]{This figure shows the result of the following survey questions:
Q1. ``Which was the easier way to enter the response, by recording the response on `paper' or by recording the response on the `Web Form (Google Spreadsheets)'?'',
Q2. ``Which was the easier way to enter the response, by recording the response on `paper' or by recording the response on `Response Collector'?'' and
Q3. ``Which was the easier way to enter the response, by recording the response on `Response Collector' or by recording the response on the `Web Form (Google Spreadsheets)'?''}{fig_q1_3}{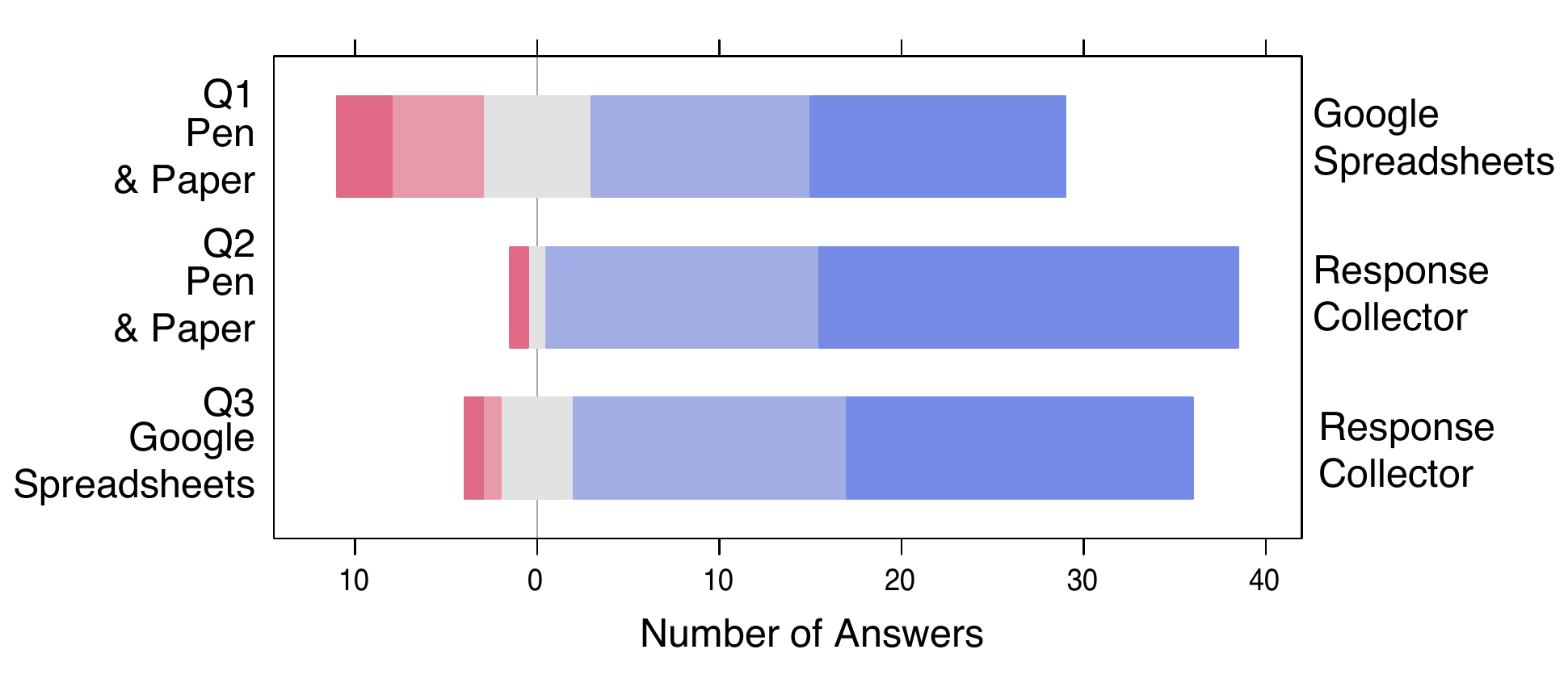}

\subsection{Input Interface Comparison with Other Methods}
This section answers RQ(B): ``Which type of inputting method do students prefer?''

\figref{fig_q1_3} shows the result of the survey question: ``For which method was it easier to input the response?''
Students chose Google Spreadsheets over pen and paper (signed test: $p < 0.01$).
They chose Response Collector over pen and paper ($p < 0.01$).
They also chose Response Collector over Google Spreadsheets ($p < 0.01$).
This result shows that it is significantly easier to input into Response Collector than other methods.

Moreover, \figref{fig_q15} shows the result of the question: ``It was easy to input responses and questions into Response Collector.''
We obtained no negative answers from students in relation to the ease of use of Response Collector.
These results lead to the conclusion that our system is easy for students to use.

\fig[width=0.98\columnwidth]{This figure shows the result of the question: Q15. ``It was easy to input responses and questions into Response Collector.'' We asked students about this twice. This result reflects that Response Collector is easy to use.}{fig_q15}{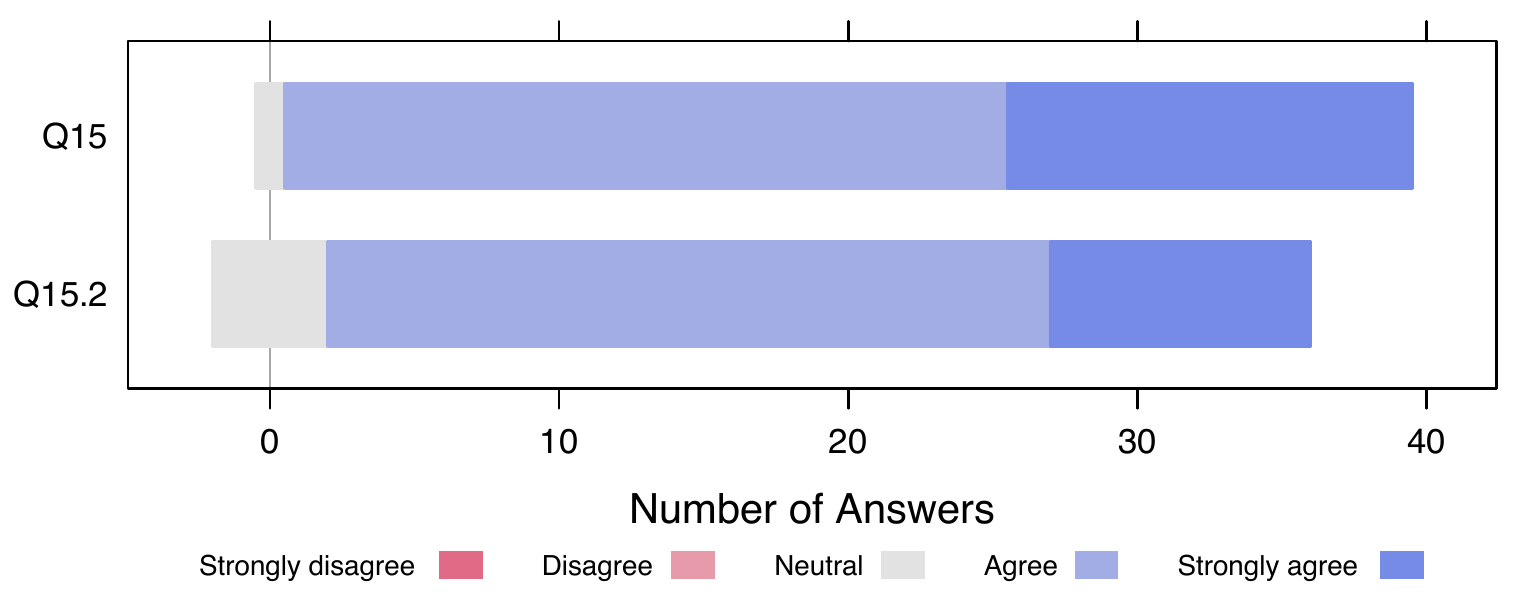}

\subsection{The Usefulness of Sharing Responses}
This section answers RQ(C): ``Is it useful for students to share information about their responses in the classroom?''

\figref{fig_q9_11} shows the answers to the survey questions that ask for a shared range where students can share their responses.
The answers to Q9 show that students disagree with limiting the range for sharing their responses to their teacher only (signed test: $p < 0.01$).
The answers to Q10 show that students tend to disagree with restricting their responses to sharing only during class, but there is no significant difference ($p < 0.09$).
The answers to Q11 show that students agree with sharing their responses freely with all students ($p < 0.01$).

\figref{fig_q26} shows the answer to the survey question: ``Can you find other observations or different points of view by viewing the responses or aggregation graph of other students using Response Collector?''
The result shows that students agreed with this ($p < 0.01$).
This indicates that Response Collector can provide different points of view by showing the responses from students.

Free descriptions of people who agree with sharing are as follows:
``Even if I do not wonder at the time of viewing, there might be things that I cannot understand when I watch later. So, it is good to see everyone's questions or what they thought important then.'' and ``By looking at points that others thought to be essential or wondered about, we may notice points that I couldn't notice.''
Free descriptions of people who disagree with sharing are as follows:
``I felt a little embarrassed'' and
``I think that it is okay to publish cases where personal information cannot be seen.
However, I feel it is unnecessary to share the responses with other students.
As I use this tool for reviewing, I think that the marks of others do not matter for me and will get in the way.''

These opinions show that students preferred to share responses with other students because they could obtain helpful ideas from others' responses that they did not notice themselves.
This indicates that they aimed to get to know others' opinions because they had confidence in themselves.
On the other hand, some students disliked sharing their own responses because they were shown with their name and student ID.
This reflects that students may prefer to remain anonymous.

\subsection{Satisfaction with the Class using Response Collector}
This section answers RQ(D): ``Are students satisfied with flipped classes where the teacher gives feedback on their responses from Response Collector?''

\figref{fig_q23_24} shows the answers to the following survey questions:
Q23. ``Did you resolve the question you recorded in Response Collector during the explanation time?'' and
Q24. ``Was it good to be able to receive explanations based on your questions in Response Collector?''
Most students answered Q23 neutrally during the first survey.
However, after completing all the classes, the students' answers shifted from ``Neutral'' to ``Agree'' (signed test: $p < 0.014$).
Moreover, the answers to Q24 shifted slightly from ``Agree'' to ``Strongly Agree.''
This result shows that the lecture assisted by Response Collector satisfied students because they could receive explanations based on their questions.

\fig[width=0.98\columnwidth]{This figure shows the result of the following survey questions:
Q9. ``Would it be better for you if only the teacher could see the responses that students recorded? (Did not disclose them to any other students at all.)'',
Q10. ``Would it be better for you if the teacher temporarily shared the responses on the screen during class to all students?'' and
Q11. ``Would it be better for you if students could see the responses freely?''}{fig_q9_11}{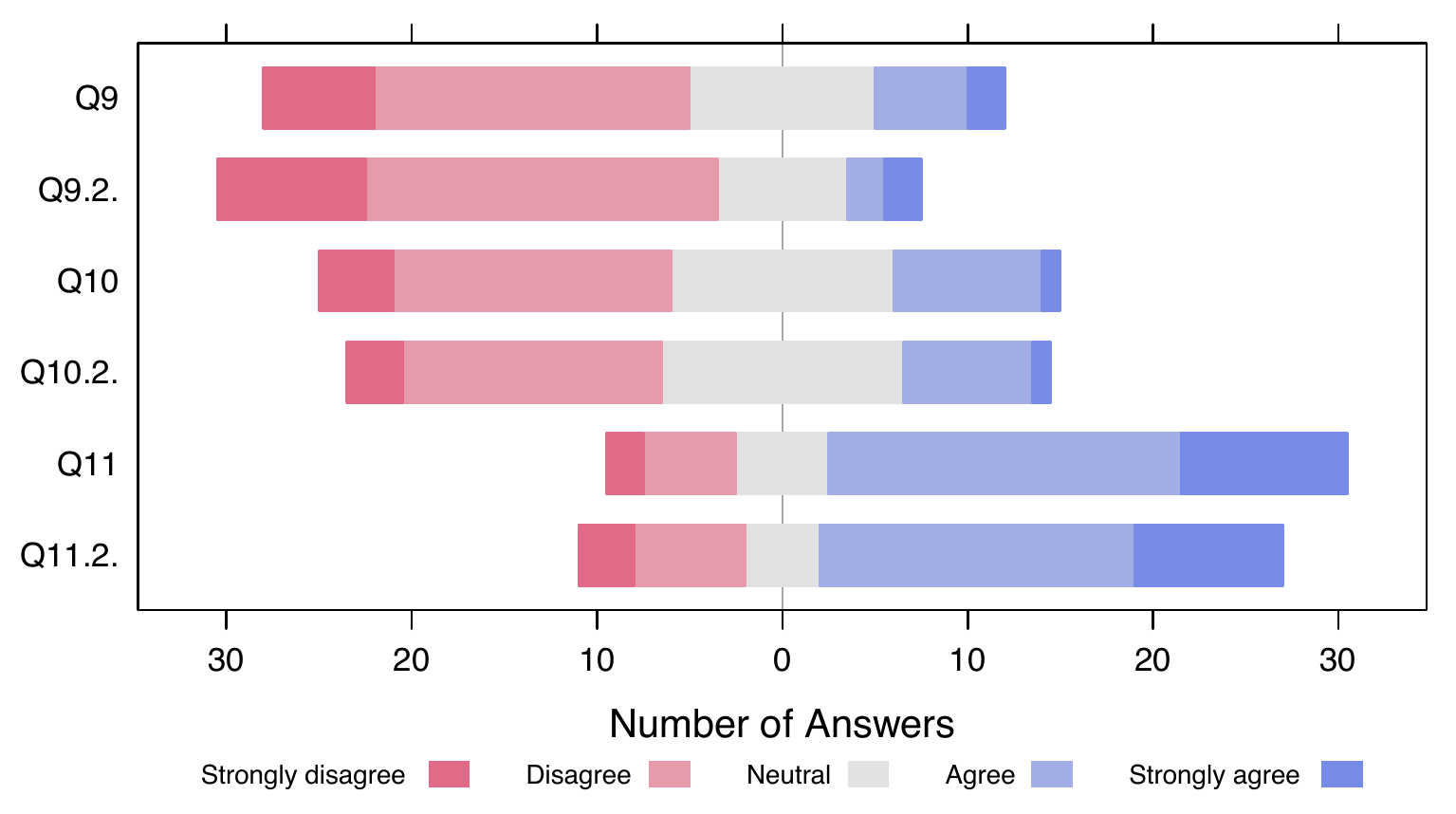}

\fig[width=0.95\columnwidth]{This figure shows the result of the survey question:
Q26. ``Can you find other observations or different points of view by viewing the responses or aggregation graphs of other students using Response Collector?''
}{fig_q26}{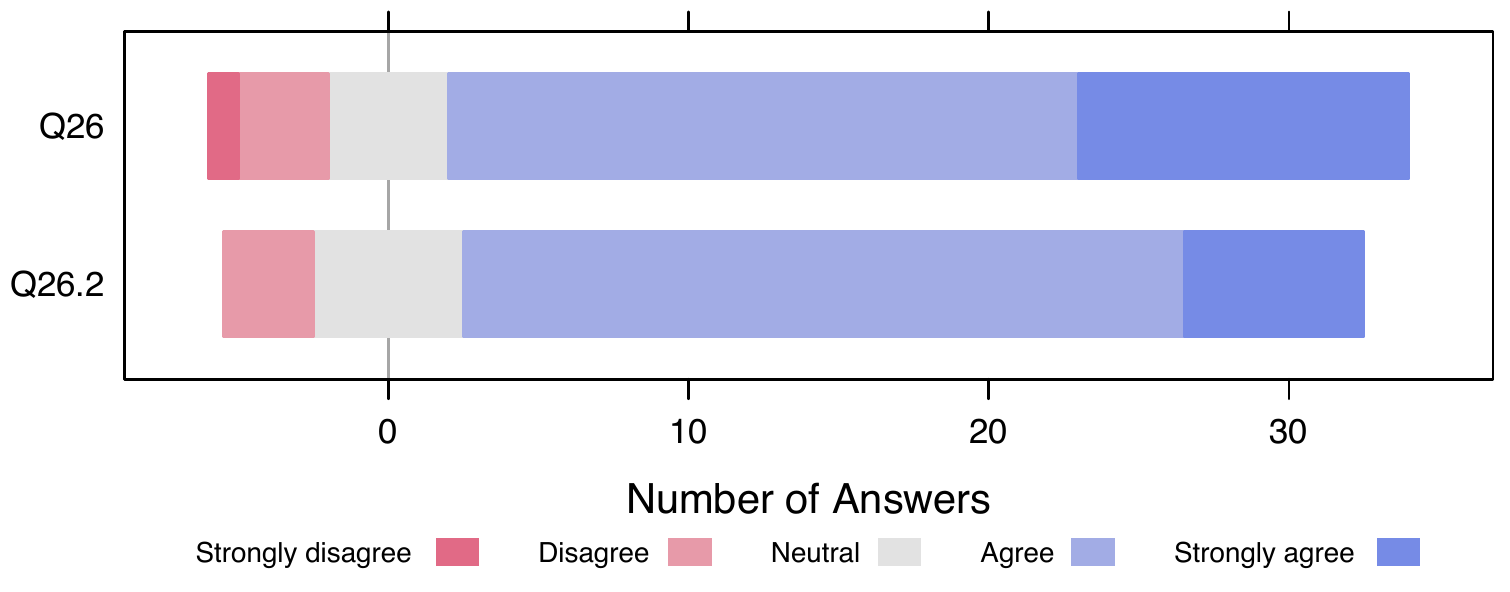}

\section{Discussion / Future Work}\label{sec_discussion}
Generally, in the flipped classroom, students watch videos before class hours.
However, it is problematic that some students do not watch videos in advance.
This paper does not focus on resolving this problem.
In our user study, we provided preparation time at the beginning of each lecture because there was sufficient time to do so.
A part of our user study also showed that some students did not watch the videos.
Therefore, we consider it better to provide students with preparation time if there is sufficient time to watch videos in class.

We were not concerned about the anonymity of the responses on Response Collector.
The user study showed that some students preferred to remain anonymous.
Thus, a future study should examine whether the responses would increase if Response Collector remained anonymous.

\fig[width=0.95\columnwidth]{This figure shows the result of the answers to the following survey questions:
Q23. ``Did you resolve the question you recorded in Response Collector during\ the explanation time?'',
Q24. ``Was it good to be able to receive explanations based on your questions in Response Collector?''
}{fig_q23_24}{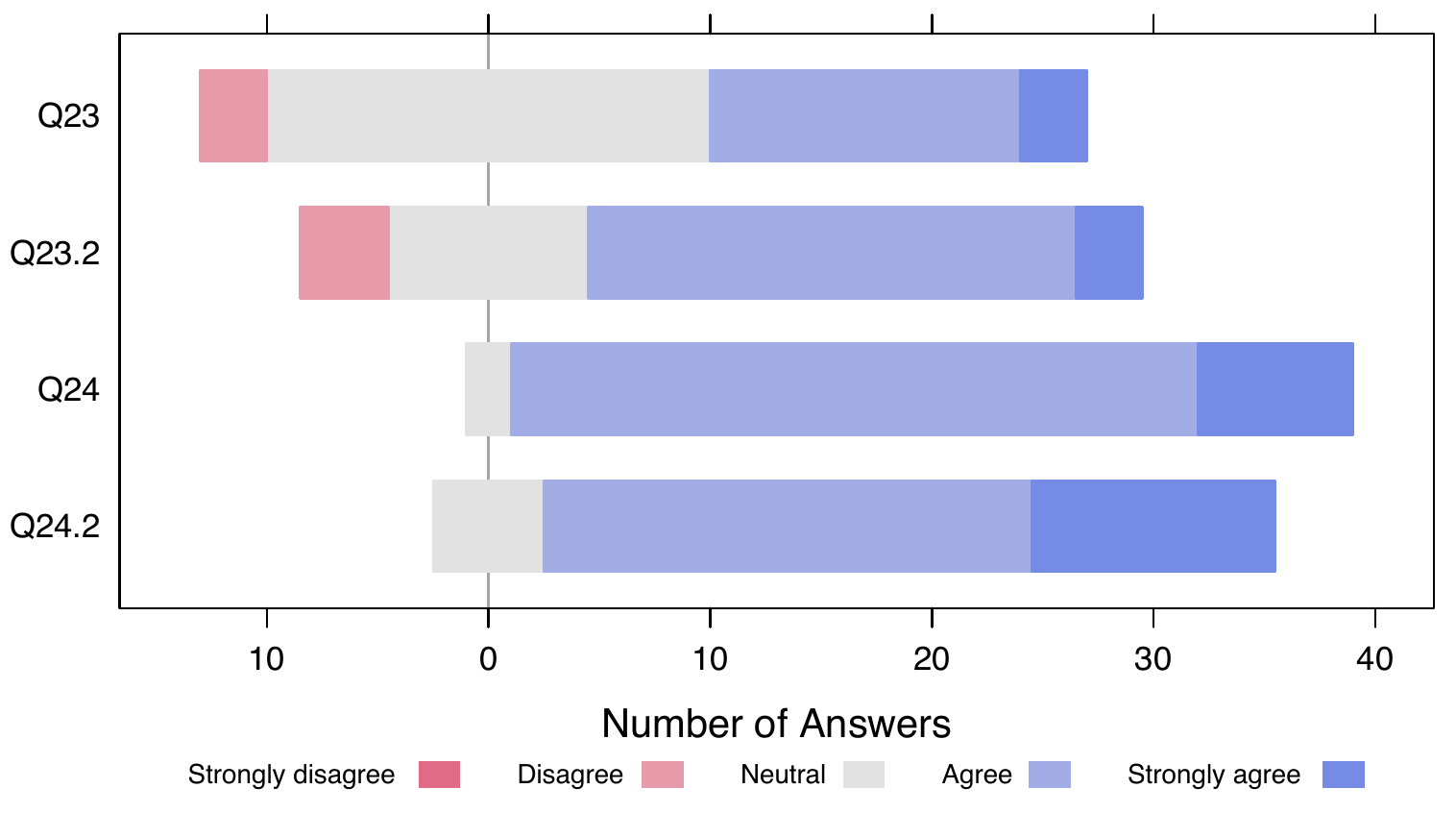}

\section{Conclusion}
In this paper, we proposed Response Collector as a video learning system for flipped classrooms, enabling teachers and students to share their responses to preparation videos.
The result of our user study showed that students preferred the proposed system as an input method over the pen and paper and Google Spreadsheets methods.
Moreover, sharing responses among students was helpful for resolving each students' questions, and students were satisfied with the use of our system in a flipped classroom setup.

\ifblind
\else
  \section*{Acknowledgment}
  A part of this research was supported by JSPS KAKENHI Grant Number 17KO0484.
\fi

\bibliographystyle{IEEEtran}

\bibliography{main}

\begin{thebibliography}{10}
\providecommand{\url}[1]{#1}
\csname url@samestyle\endcsname
\providecommand{\newblock}{\relax}
\providecommand{\bibinfo}[2]{#2}
\providecommand{\BIBentrySTDinterwordspacing}{\spaceskip=0pt\relax}
\providecommand{\BIBentryALTinterwordstretchfactor}{4}
\providecommand{\BIBentryALTinterwordspacing}{\spaceskip=\fontdimen2\font plus
\BIBentryALTinterwordstretchfactor\fontdimen3\font minus
  \fontdimen4\font\relax}
\providecommand{\BIBforeignlanguage}[2]{{%
\expandafter\ifx\csname l@#1\endcsname\relax
\typeout{** WARNING: IEEEtran.bst: No hyphenation pattern has been}%
\typeout{** loaded for the language `#1'. Using the pattern for}%
\typeout{** the default language instead.}%
\else
\language=\csname l@#1\endcsname
\fi
#2}}
\providecommand{\BIBdecl}{\relax}
\BIBdecl

\bibitem{Bergmann2012}
J.~Bergmann and A.~Sams, \emph{{Flip Your Classroom: Reach Every Student in
  Every Class Every Day}}.\hskip 1em plus 0.5em minus 0.4em\relax ISTE, 2012.

\bibitem{Bishop2013}
J.~L. Bishop and M.~A. Verleger, ``{The flipped classroom: A survey of the
  research},'' in \emph{Proc. ASEE National Conference}, 2013, pp. 1--18.

\bibitem{Thai2017}
N.~T.~T. Thai, B.~{De Wever}, and M.~Valcke, ``{The impact of a flipped
  classroom design on learning performance in higher education: Looking for the
  best “blend” of lectures and guiding questions with feedback},'' \emph{J.
  Computers {\&} Education}, vol. 107, pp. 113--126, Apr 2017.

\bibitem{Bargeron1999}
D.~Bargeron, A.~Gupta, J.~Grudin, and E.~Sanocki, ``{Annotations for streaming
  video on the Web: system design and usage studies},'' \emph{J. Computer
  Networks}, vol.~31, no. 11-16, pp. 1139--1153, May 1999.

\bibitem{Bargeron2002}
D.~Bargeron, J.~Grudin, A.~Gupta, E.~Sanocki, F.~Li, and S.~{Lee Tiernan},
  ``{Asynchronous Collaboration Around Multimedia Applied to On-Demand
  Education},'' \emph{J. Management Information Systems}, no.~4, pp. 117--145,
  Mar 2002.

\bibitem{Chatti2016}
M.~A. Chatti, M.~Marinov, O.~Sabov, R.~Laksono, Z.~Sofyan, A.~M. {Fahmy
  Yousef}, and U.~Schroeder, ``{Video annotation and analytics in
  CourseMapper},'' \emph{J. Smart Learning Environments}, vol.~3, no.~1, p.~10,
  Dec 2016.

\bibitem{Risko2013}
E.~F. Risko, T.~Foulsham, S.~Dawson, and A.~Kingstone, ``{The Collaborative
  Lecture Annotation System (CLAS): A New TOOL for Distributed Learning},''
  \emph{IEEE Trans. Learning Technologies}, vol.~6, no.~1, pp. 4--13, Jan 2013.

\bibitem{Mirriahi2016}
N.~Mirriahi, D.~Liaqat, S.~Dawson, and D.~Ga{\v{s}}evi{\'{c}}, ``{Uncovering
  student learning profiles with a video annotation tool: reflective learning
  with and without instructional norms},'' \emph{J. Educational Technology
  Research and Development}, vol.~64, no.~6, pp. 1083--1106, Dec 2016.

\bibitem{Kim2014}
J.~Kim, P.~J. Guo, D.~T. Seaton, P.~Mitros, K.~Z. Gajos, and R.~C. Miller,
  ``{Understanding In-video Dropouts and Interaction Peaks In online Lecture
  Videos},'' in \emph{Proc. the 1st ACM Conference on Learning @ Scale}, 2014,
  pp. 31--40.

\bibitem{Shi2015}
C.~Shi, S.~Fu, Q.~Chen, and H.~Qu, ``{VisMOOC: Visualizing video clickstream
  data from Massive Open Online Courses},'' in \emph{Proc. 2015 IEEE Pacific
  Visualization Symposium}, 2015, pp. 159--166.

\end{thebibliography}

\end{document}